\documentclass[english]{svjour3}
\usepackage[T1]{fontenc}
\usepackage[latin9]{inputenc}
\usepackage{wrapfig}
\usepackage{mathrsfs}
\usepackage{url}
\usepackage{amsmath}
\usepackage{amssymb}
\usepackage{graphicx}
\usepackage{wasysym}
\usepackage{esint}
\usepackage{subscript}

\makeatletter

\providecommand{\tabularnewline}{\\}

\RequirePackage{fix-cm}

\smartqed  % flush right qed marks, e.g. at end of proof
\usepackage{physics}

\makeatother

\usepackage{babel}
\begin{document}

\title{Thermal conductivity of superfluid \textsuperscript{3}He-B in a
tubular channel down to 0.1\emph{T}\textsubscript{c} at the \textsuperscript{4}He
crystallization pressure\thanks{}}

\titlerunning{Thermal conductivity of superfluid \textsuperscript{3}He-B}

\author{T. S. Riekki \and J. T. Tuoriniemi \and A. P. Sebedash }

\authorrunning{T. S. Riekki \and  J. T. Tuoriniemi\and  A. P. Sebedash }

\institute{T. S. Riekki\and  J. T. Tuoriniemi\at Aalto University\\
School of Science\\
Low Temperature Laboratory\\
P.O. BOX 15100 FI-00076 Aalto, Finland\\
\email{tapio.riekki@aalto.fi}\\
A. P. Sebedash \at P. L. Kapitza Institute for Physical Problems
RAS\\
Kosygina 2, 119334 Moscow, Russia\\
}

\date{Received: date / Accepted: date}
\maketitle
\begin{abstract}
We studied the thermal conductivity of superfluid \textsuperscript{3}He in a 2.5~mm effective diameter and 0.15~m long channel connecting the two volumes of our experimental assembly. The main volume contained pure solid \textsuperscript{4}He, pure liquid \textsuperscript{3}He and saturated liquid \textsuperscript{3}He--\textsuperscript{4}He mixture at varying proportions, while the separate heat-exchanger volume housed sinter and was filled by liquid \textsuperscript{3}He. The system was cooled externally by a copper nuclear demagnetization stage, and, as an option, internally by the adiabatic melting of solid \textsuperscript{4}He in the main volume. The counterflow effect of superfluid just below the transition temperature $T_c$ resulted in the highest observed conductivity about five times larger than that of the normal fluid at the $T_c$. Once the hydrodynamic contribution had practically vanished below $0.5T_c$, we first observed almost constant conductivity nearly equal to the normal fluid value at the $T_c$. Finally, below about $0.3T_c$, the conductivity rapidly falls off towards lower temperatures.

\keywords{Helium-3\and Helium-4\and Helium-3\textendash Helium-4 mixture \and superfluid
thermal conductivity}
\end{abstract}

\section{Introduction\label{sec:Introduction}}

Thermal conductivity of superfluid \textsuperscript{3}He consists
of two components: diffusive conductivity due to the quasiparticle
motion, and hydrodynamic conductivity caused by the superfluid-normal
fluid counterflow effect \cite{London1948}. The hydrodynamic conductivity
is most important just below the superfluid transition temperature
$T_{c}$, as it requires the presence of the normal component, whose amount decreases exponentially with temperature. Diffusive conductivity
has been discussed in a few theoretical publications \cite{Pethick1975,Pethick1977,Dorfle1980,Hara1981},
and has been measured using a heat-pulse method \cite{Wellard1982,Einzel1984}.
Measurements of the total thermal conductivity have
been made only on a narrow temperature span near the $T_{c}$ \cite{Greytak1973,Johnson1975} at a selection of pressures, and at a single point at the \textsuperscript{3}He crystallization pressure in the ballistic quasiparticle regime \cite{Feng1993}.

Our interest in the matter is related to our adiabatic melting experiment
that aims to cool \textsuperscript{3}He and saturated \textsuperscript{3}He\textendash \textsuperscript{4}He
mixture to ultra-low temperatures at the \textsuperscript{4}He crystallization
pressure 2.564 MPa \cite{Adiabatic_Melting,Sebedash_QFS}. The method is capable
of reaching temperatures below $0.1\,\mathrm{mK}$ by melting solid
\textsuperscript{4}He and mixing it with liquid \textsuperscript{3}He.
At the lowest achievable temperatures, our quartz tuning fork thermometers
become insensitive \cite{Riekki2019a}, and a computational modeling
of the system is required to evaluate the temperature. To carry out the simulation, we need good understanding of the thermal couplings within
the system, of which one of the key components is the thermal conductivity of superfluid \textsuperscript{3}He. Also, the thermal boundary resistance between liquid helium and the
cell wall, and between liquid and the sintered heat-exchanger are needed.

Our experimental setup provides a unique opportunity to map such intricate
thermal parameters across a wide temperature range at various thermal
loads, as the total heat capacity of the system can be varied
by altering the amount of mixture in the system by changing the size
of the \textsuperscript{4}He crystal. Temperatures from $10\,\mathrm{mK}$
down to $0.5\,\mathrm{mK}$ were reached by cooling the system by
a nuclear demagnetization refrigerator, while temperatures below that
were accessible by the adiabatic melting method. We were thus able
to study the thermal conductivity of \textsuperscript{3}He down to
the low temperature limit of our thermometry.

\section{Thermal model}

\begin{figure}[t]
	\centering
	\includegraphics[width=.75\columnwidth]{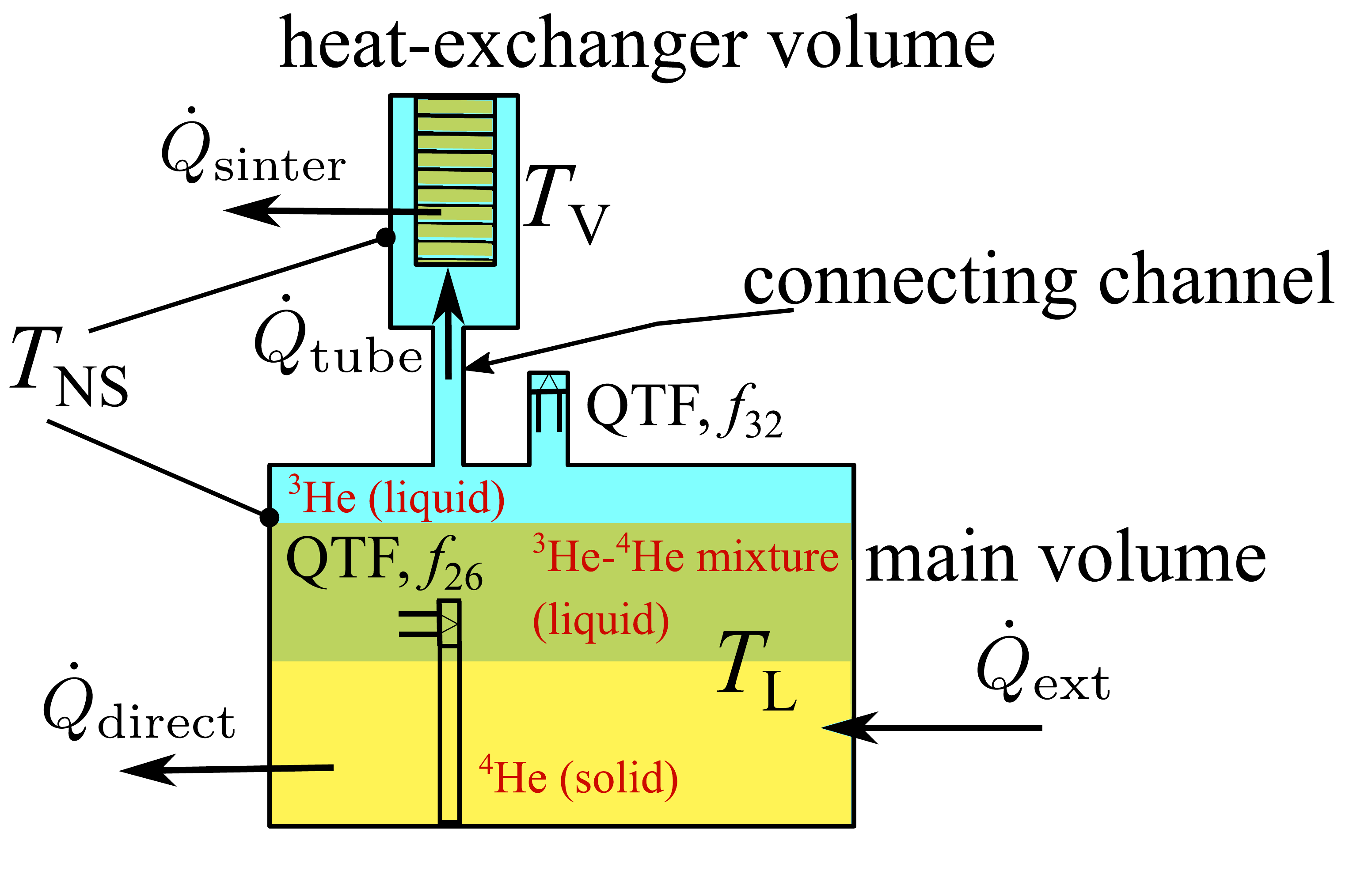}\caption{(color online) Simplified drawing of the experimental cell, showing phases,
		heat flows and temperatures in the system during precool. Liquid in the main volume
		is at $T_{\mathrm{L}}$, in the heat-exchanger volume at $T_{\mathrm{V}}$,
		while all container walls are thermalized to the precooler (nuclear
		stage) at $T_{\mathrm{NS}}$. The main thermometer QTF ($f_{32}$)
		is located on the top section of the main volume\label{fig:cell-drawing} }
\end{figure}

The experimental cell, shown schematically in Fig.~\ref{fig:cell-drawing},
consisted of a main volume ($77\,\mathrm{cm^{3}}$), and a sinter-filled
heat-exchanger volume ($5\,\mathrm{cm^{3}}$), connected together
by a tubular channel \cite{Sebedash_QFS}. The main volume was filled with pure solid \textsuperscript{4}He,
saturated liquid \textsuperscript{3}He\textendash \textsuperscript{4}He mixture
and pure liquid \textsuperscript{3}He at varying proportions, whereas
the heat-exchanger volume and the connecting channel were always filled by
pure \textsuperscript{3}He. There may have also been small amount of
mixture trapped in the porous sinter. The main volume was monitored
by two quartz tuning fork oscillators (QTFs): one situated in an extension at the top of the main volume to always keep it in the pure \textsuperscript{3}He phase, while the other was in the middle
of the main volume, and thus in the mixture phase, or frozen in solid
\textsuperscript{4}He, depending on the size of the \textsuperscript{4}He
crystal. Additionally, our setup had a cold
valve (not shown in Fig.~\ref{fig:cell-drawing}) that could be used
to restrict the channel, but it was kept open during the
measurements described here. The cell was precooled by copper nuclear adiabatic demagnetization cooler, whose temperature was measured by a pulsed \textsuperscript{195}Pt NMR thermometer.

There were two filling
lines to the cell: a normal capillary attached to the heat-exchanger volume, and a
superleak line to the main volume. The normal capillary was used to introduce \textsuperscript{3}He to the system, but after that it was blocked by solid helium, while the superleak line was
usually open to a reservoir at about $10\,\mathrm{mK}$ temperature,
and to Kelvin-range environment from there on. The superleak line was a capillary filled with tightly packed metal-oxide powder, whose large impedance allows only superfluid flow. The crystallization pressure in porous materials is larger than in bulk, which is why the superleak can be used to transfer \textsuperscript{4}He to (or from) the cell at the bulk crystallization pressure to grow (or melt) the solid phase. 

We have a univariant three-phase system, as solid \textsuperscript{4}He fixes the pressure to its crystallization
pressure $2.564\,\mathrm{MPa}$ \cite{Pentti_Thermometry}, while the
presence of the pure \textsuperscript{3}He phase ensures that the
mixture is at its saturation concentration $8.12\%$ \cite{Pentti_etal_solubility},
leaving only temperature $T$ as a free variable. When the cell is
cooled down, \textsuperscript{3}He becomes A-phase superfluid at
$T=T_{c}=2.6\,\mathrm{mK}$ \cite{Pentti_Thermometry}, while the transition
to the B-phase occurs at $T_{AB}=0.917T_{c}$ \cite{Greywall1986}$\approx2.4\,\mathrm{mK}$.

During external cooling, heat flows from the main volume of the experimental
cell ($T_{\mathrm{L}}$) to the precooler ($T_{\mathrm{NS}}$) via two
paths: directly through the plain cell wall, which becomes rather unimportant below the $T_c$, and through the sinter of
the heat-exchanger volume via the connecting channel. Due to better
thermal connection to the precooler, the liquid in the heat-exchanger volume ($T_{\mathrm{V}}$) follows temperature
changes faster, while the main volume lags behind. This gives us an opportunity to evaluate the thermal conductivity of superfluid \textsuperscript{3}He in the connecting channel.

The Kapitza resistance
$R_{K}$ from liquid helium to the cell wall and to the sinter are assumed to obey a
power law
\begin{equation}
R_{K}=\frac{R_{0}}{AT^{p}},\label{eq:kapitza}
\end{equation}
where $A$ is the surface area, while $R_{0}$ and $p$ are constants that depend on the temperature range and materials in question, as
discussed in Refs.~\cite{Voncken_sinter,Busch1984,Oh1994,Castelijns1985,Riekki_long}. We measured the sinter to have approximately
$10\,\mathrm{m^{2}}$ surface area, while the cell wall area was estimated to be $0.12\,\mathrm{m^{2}}$.
In the following treatment we combine $R_{0}$ and $A$ into one
parameter $r=A/R_{0}$. Thus, the heat flow across the Kapitza resistance
becomes
\begin{equation}
\dot{Q}_{K}\left(p,r,T\right)=\intop_{T}^{T_{\mathrm{NS}}}\frac{\mathrm{d}T'}{R_{K}}=\frac{r}{p+1}\left(T_{\mathrm{NS}}^{p+1}-T^{p+1}\right),\label{eq:kapitza-heat}
\end{equation}
where $r$ and $p$ have different values for the sinter and the plain
cell wall. 

Pure \textsuperscript{3}He phase and the mixture phase in the main volume (L) are assumed to have uniform temperature, and thus the heat balance there reads
\begin{equation}
C_{\mathrm{L}}\dot{T}_{\mathrm{L}}\left(t\right)=\dot{Q}_{\mathrm{melt}}+\dot{Q}_{\mathrm{ext}}+\dot{Q}_f+\dot{Q}_{\mathrm{direct}}+\dot{Q}_{\mathrm{tube}}.\label{eq:L-heat}
\end{equation}
A dot above a symbol indicates derivative with respect to time $t$. Here $C_{\mathrm{L}}=n_{3}^{\mathrm{L}}C_{3}+n_{{m,3}}^{\mathrm{L}}C_{{m,3}}$
is the heat capacity of the liquid in the main volume, with $n_{3}^{\mathrm{L}}$ and $n_{{m,3}}^{\mathrm{L}}$
the amounts of \textsuperscript{3}He in the pure \textsuperscript{3}He
phase and in the saturated \textsuperscript{3}He\textendash \textsuperscript{4}He
mixture phase, respectively, while $C_{3}$ and $C_{{m,3}}$
are their heat capacities per mole of \textsuperscript{3}He, respectively. The first term on the right side is
the heat absorbed (or released) when solid \textsuperscript{4}He
is melted (or grown), $\dot{Q}_{\mathrm{melt}}=T_{\mathrm{L}}\dot{n}_{\mathrm{3}}\left(S_{{m,3}}-S_{3}\right)$ \cite{Riekki2019},
where $\dot{n}_{\mathrm{3}}$ is the rate at which \textsuperscript{3}He
is transferred between the liquid phases, and $S_{3}$ and $S_{{m,3}}$
are the entropies of pure \textsuperscript{3}He and saturated \textsuperscript{3}He\textendash \textsuperscript{4}He
mixture per mole of \textsuperscript{3}He, respectively. Each $C_{3}$,
$C_{{m,3}}$, $S_{3}$ and $S_{{m,3}}$ is a function
of temperature, and they are as given by Ref.~\cite{Riekki2019}
(we assume that the heat capacity of pure solid \textsuperscript{4}He
is negligibly small). Next, $\dot{Q}_{\mathrm{ext}}$
is the background heat leak to the main volume, while $\dot{Q}_f$
represents losses occurring when there is flow through the superleak line. Lastly, $\dot{Q}_{\mathrm{direct}}=\dot{Q}_{K}\left(p_{\mathrm{L}},r_{\mathrm{L}},T_{\mathrm{L}}\left(t\right)\right)$ is the heat flowing to the precooler through the plain wall Kapitza
bottleneck, given by Eq.~\eqref{eq:kapitza-heat}, and $\dot{Q}_{\mathrm{tube}}$ is the heat leaving the main volume through the connecting channel
for the heat-exchanger volume. During the precooling period, both $\dot{Q}_{\mathrm{melt}}$
and $\dot{Q}_f$ can be omitted, as solid \textsuperscript{4}He
is neither grown nor melted. This applies to all cases considered in this paper.

Next, for the heat balance of the heat-exchanger volume (V), we get

\begin{equation}
C_{\mathrm{V}}\dot{T}_{\mathrm{V}}\left(t\right)=\dot{Q}_{\mathrm{sinter}}-\dot{Q}_{\mathrm{tube}},\label{eq:V-heat}
\end{equation}
where $C_{\mathrm{V}}=n_{3}^{\mathrm{V}}C_{3}+n_{{m,3}}^{\mathrm{V}}C_{{m,3}}$, is similar to the first term of Eq.~\eqref{eq:L-heat}, $\dot{Q}_{\mathrm{sinter}}=\dot{Q}_{K}\left(p_{\mathrm{V}},r_{\mathrm{V}},T_{\mathrm{V}}\left(t\right)\right)$ is the heat flowing to the precooler through the sinter Kapitza resistance,
again given by Eq. \eqref{eq:kapitza-heat}, and $\dot{Q}_{\mathrm{tube}}$ is
the heat coming from the main volume through the channel. Here $n_{3}^{\mathrm{V}}$ includes liquid \textsuperscript{3}He in the connecting channel as well, but assumes, for simplicity, that the entire channel is at the same temperature as the heat-exchanger volume. Having the channel to be at $\left(T_{\mathrm{V}}+T_{\mathrm{L}}\right)/2$ would not modify the simulations notably, since the heat capacity of the small \textsuperscript{3}He amount in the channel is insignificant next to the heat capacity of the heat-exchanger volume. Note
that $n_{{m,3}}^{\mathrm{V}}$ can be non-zero due to the mixture
trapped into the sinter. We estimated that the sinter can hold maximum of 7 mmol of \textsuperscript{3}He. 

All the heat that is not transmitted through the cell wall must
flow through the connecting channel to the heat-exchanger volume and then
through the sinter to the precooler. The thermal resistance of the
cylindrical channel is
\begin{equation}
R_{T}=\frac{4l}{\kappa\left(T\right)\pi d^{2}},\label{eq:tube-resistance}
\end{equation}
where $\kappa\left(T\right)$ is the thermal conductivity of pure
\textsuperscript{3}He, $l\approx15\,\mathrm{cm}$, and $d\approx2.5\,\mathrm{mm}$
are the length and the effective diameter of the channel. In reality,
the channel is not equally wide along its complete length, and thus
5--10\% uncertainty in $D=\frac{\pi d^{2}}{4l}$ results. The
heat flow through such channel is given by the integral
\begin{equation}
\dot{Q}_{\mathrm{tube}}=\intop_{T_{\mathrm{V}}}^{T_{\mathrm{L}}}\frac{dT'}{R_{T}}=D\intop_{T_{\mathrm{V}}}^{T_{\mathrm{L}}}\kappa\left(T'\right)\mathrm{d}T'.\label{eq:tube-heat0}
\end{equation}

In the normal state of pure \textsuperscript{3}He, from $T=T_{c}=2.6\,\mathrm{mK}$ \cite{Pentti_Thermometry},
up to our range of interest ($T \approx 10\,\mathrm{mK}$), its thermal conductivity
follows $\kappa\left(T\right)=\kappa_{0}/T$ dependence, with the
coefficient $\kappa_{0}=9.69\cdot10^{-5}\,\mathrm{\frac{W}{m}}$ interpolated
from the data of Ref.~\cite{Greywall1984}. But below the $T_{\mathrm{c}}$,
the situation becomes more complicated, as the behavior of $\kappa\left(T\right)$
is not well established. We can proceed by first dividing the heat
flow integral of Eq.~\eqref{eq:tube-heat0} into above and below
the $T_{\mathrm{c}}$ parts, and then linearizing it below the $T_{\mathrm{c}}$.
This is a valid course of action as long as the temperature of the
heat-exchanger volume $T_{\mathrm{V}}$ does not drop far below the $T_{\mathrm{c}}$
until the main volume temperature $T_{\mathrm{L}}$ is there as well.
The integral of Eq.~\eqref{eq:tube-heat0} may thus be written as

\begin{equation}
\begin{aligned}\dot{Q}_{\mathrm{tube}} & =D\intop_{\max\left(T_{\mathrm{V}}\left(t\right),T_{c}\right)}^{\max\left(T_{\mathrm{L}}\left(t\right),T_{c}\right)}\frac{\kappa_{0}}{T'}\mathrm{d}T'+D\intop_{\min\left(T_{\mathrm{V}}\left(t\right),T_{c}\right)}^{\min\left(T_{\mathrm{L}}\left(t\right),T_{c}\right)}\kappa\left(T'\right)\mathrm{d}T'\\
 & =D\kappa_{0}\ln\left[\frac{\max\left(T_{\mathrm{L}}\left(t\right),T_{c}\right)}{\max\left(T_{\mathrm{V}}\left(t\right),T_{c}\right)}\right]+D\kappa_{1}\left(T_{\mathrm{tube}}\right)\left[\min\left(T_{\mathrm{L}}\left(t\right),T_{c}\right)-\min\left(T_{\mathrm{V}}\left(t\right),T_{c}\right)\right],
\end{aligned}
\label{eq:tube-heat}
\end{equation}
where $\kappa_{1}\left(T_{\mathrm{tube}}\right)$ is the superfluid \textsuperscript{3}He thermal
conductivity at the average channel temperature $T_{\mathrm{tube}}$.

\section{Results\label{sec:Results}}

The results presented here were obtained by analyzing 8 precools from
about $10\,\mathrm{mK}$ ($\approx4T_{c}$) to $0.5\,\mathrm{mK}$
($\approx0.2T_{c}$), 9 low temperature precools between $1.5\,\mathrm{mK}$
($\approx0.6T_{c}$) and $0.5\,\mathrm{mK}$, as well as 5 warm-up
periods after melting of solid \textsuperscript{4}He at temperatures
below $0.5\,\mathrm{mK}$. The thermal transport parameters presented
here were determined so that all those precools and warm-ups could
be computationally reproduced within reasonable accuracy.

The challenge is that the three heat conduction paths, direct
conduction through the plain cell wall, conduction through the connecting
channel, and conduction through the sinter are intertwined, hence
none of them can be determined truly independently. Fortunately, certain
stages of the precool are more sensitive to one than the others. At
the beginning of the precool, the temperature is so high that heat
conduction through the surface of the main cell volume brings along
a significant contribution to the total heat transfer, even if the surface of the cell is hundred
times less than the surface area of the sinter. On the other hand,
thermal conductivity of \textsuperscript{3}He in the connecting channel
plays important role near the $T_{c}$ of the main cell volume, as
its conductivity increases significantly due to the superfluid-normal
fluid counterflow effect. The Kapitza resistance of the sinter is
somewhat difficult to discern, since it contributes to the heat flow
over the entire temperature range. However, it is effectively decoupled from
the main volume due to the relatively poor thermal conductivity along
the connecting channel at temperatures well above the $T_{c}$. The
path through the channel and the sinter is overwhelmingly dominant
anywhere below the $T_{c}$, which was obviously intended, as the
sole purpose of the sinter was to enable precooling the experimental
cell to as far below $1\,\mathrm{mK}$ as possible.

\begin{figure}
	\centering
	\includegraphics[width=1\columnwidth]{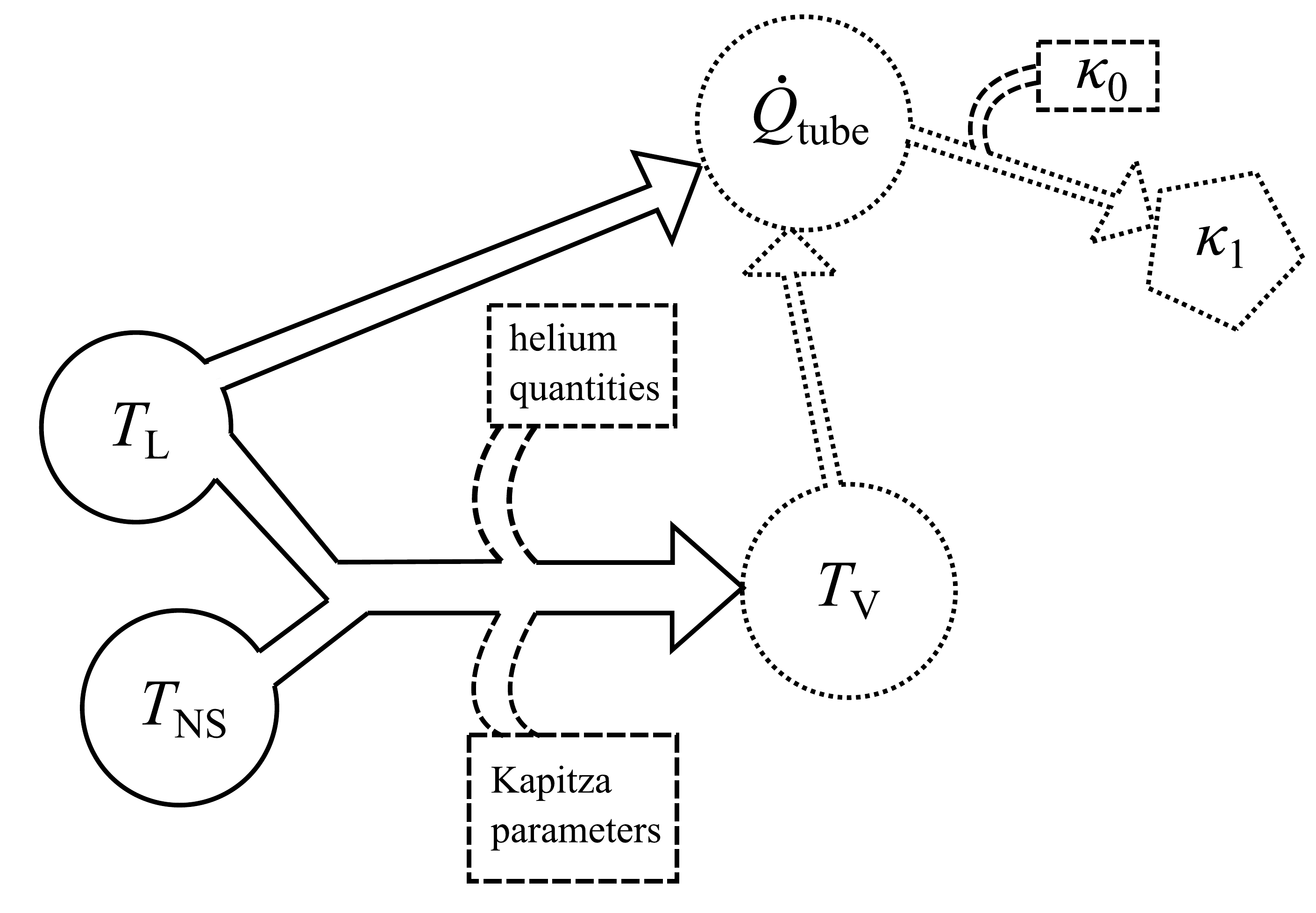}
	\caption{Flowchart of the analysis procedure. Measured quantities are indicated by solid outline, computed quantities by dotted lines, and quantities obtained by other means by dashed lines ($\kappa_{0}$ from Ref.~\cite{Greywall1984}, helium quantities from our log, and Kapitza parameters from Ref.~\cite{Riekki_long})}\label{fig:flowchart}
\end{figure}

Further challenge is provided by the varying background heat leak $\dot{Q}_{\mathrm{ext}}$
to the main cell volume. This is mostly consequential at temperatures
below 1 mK. We observed that it depended on whether the cold valve
was filled with liquid helium or not, and if there had been flow through
the superleak recently. In our analysis we have let it vary from $20\,\mathrm{pW}$
to $300\,\mathrm{pW}$ to make computations match with the experimental
observations. The highest heat leak occurred when the magnetic field
of the nuclear demagnetization stage was changing, while the maximum
idle state heat leak was about $80\,\mathrm{pW}$. For each precool and post-melting warm-up period, we used a constant heat leak value.

The procedure used to resolve the thermal conductivity
of \textsuperscript{3}He below the $T_{c}$ is illustrated in Fig.~\ref{fig:flowchart}. First, we solve differential
Eqs. \eqref{eq:L-heat} and \eqref{eq:V-heat} for $T\mathrm{_{V}}\left(t\right)$,
as the main volume temperature $T\mathrm{_{L}}$ is known based on the QTF measurements, and the nuclear stage temperature $T\mathrm{_{NS}}$ from the PLM measurement. Throughout the measurement we kept a log of the amount of helium in the different phases to calculate the heat capacities at each stage. For the plain cell wall Kapitza resistance,
we used the values $r_{\mathrm{L}}=0.69\,\mathrm{W\,K^{-3.6}}$ and $p_{\mathrm{L}}=2.6$,
determined by analyzing the precooling data near $10\,\mathrm{mK}$.
The heat-exchanger volume Kapitza parameters on the other hand were $r_{\mathrm{V}}=0.18\,\mathrm{W\,K^{-2.7}}$
and $p_{\mathrm{V}}=1.7$, which were determined at temperatures
below $2\,\mathrm{mK}$ by repeatedly growing
or melting small amount of solid \textsuperscript{4}He to alter the heat capacity of the system, and studying how that changed the relaxation time of the system toward the precooler temperature. The detailed account of that
analysis can be found in Ref.~\cite{Riekki_long}.

\begin{figure}[t]
\includegraphics[width=1\columnwidth]{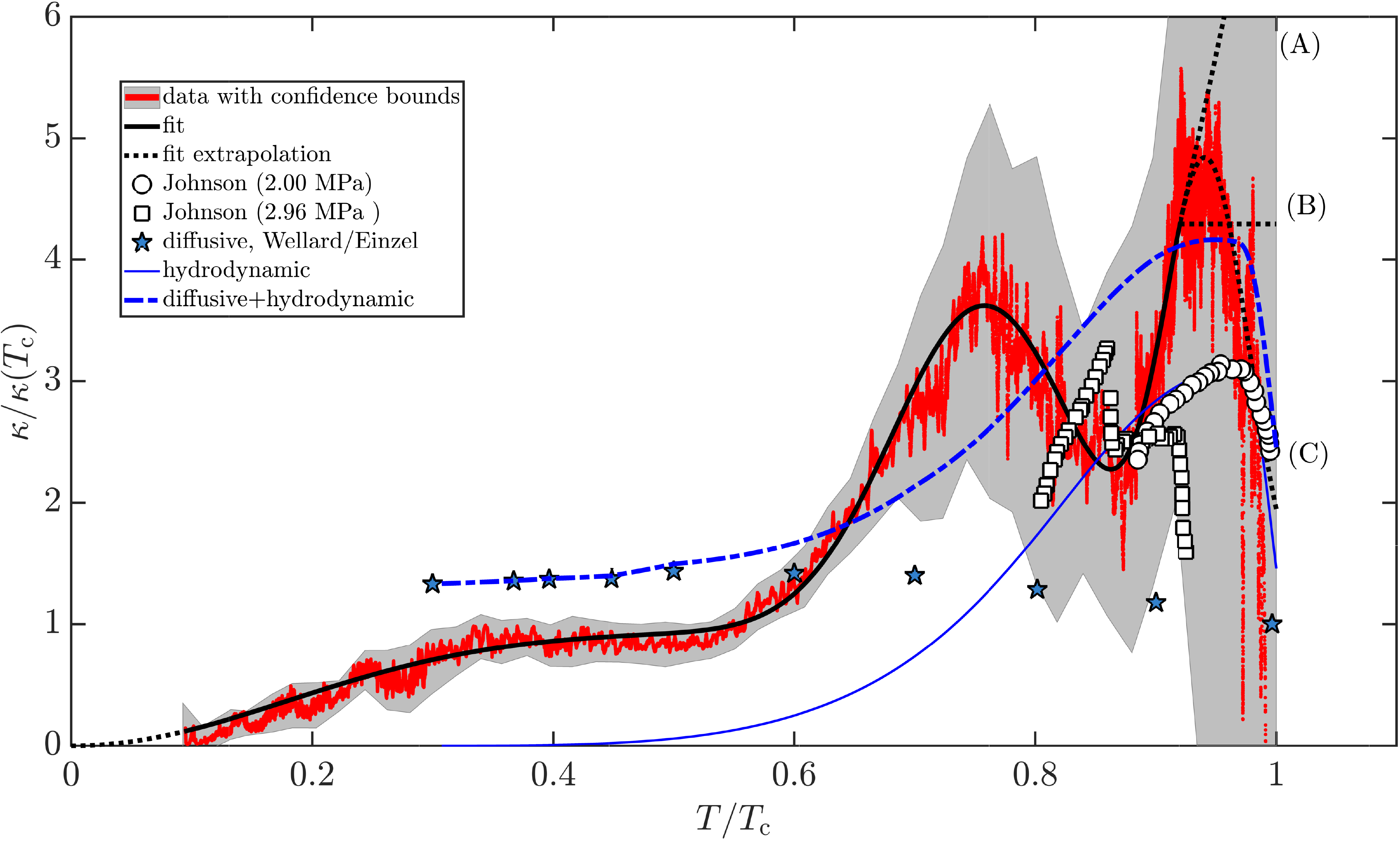}

\caption{(color online) Thermal conductivity of superfluid \protect\textsuperscript{3}He
at $2.564\,\mathrm{MPa}$ as a function of temperature relative to
the superfluid transition temperature $T_{c}$ is shown in red with
the shaded gray area representing the confidence bounds. Thermal conductivity
values are scaled by the normal fluid value at the $T_{c}$ ($0.037\,\mathrm{\tfrac{W}{K\,m}}$ \cite{Greywall1984}).
The solid black line is a multi-Gaussian fit  $G\left(T\right)$ to the data, and the
dotted black lines show the extrapolation of the fit at $T<0.1T_{c}$
and at $T>0.92T_{c}$. From $0.92T_{c}$ to the $T_{c}$, three possible
options are given: (A) linear increase from $4.3$ to $8.0\,\tfrac{\kappa}{\kappa(T_{c})}$,
(B) constant $4.3\,\tfrac{\kappa}{\kappa(T_{c})}$, and (C)  $G\left(T\right)$ up to the $T_{c}$. Thermal conductivity data by Johnson \emph{et
al}. \cite{Johnson1975} at $2.00\,\mathrm{MPa}$ ($\Circle$) and
at $2.96\,\mathrm{MPa}$ ($\square$), scaled by the normal fluid conductivity at the $T_c$ for each pressure, alongside with diffusive thermal
conductivity ($\bigstar$) by Wellard \emph{et al.} \cite{Wellard1982} (further analyzed by Einzel \cite{Einzel1984}),
as well as hydrodynamic conductivity (solid blue line) calculated
from Eq.~\eqref{eq:hydro_conductivity} are shown for comparison.
The dash-dotted blue line shows the diffusive and hydrodynamic conductivities
combined}\label{fig:thermal-conductivity}
\end{figure}

Heat transmitted through the channel, as evaluated from Eq.~\eqref{eq:L-heat}, depends on the derivative $\mathrm{d}/\mathrm{d}t$
of the liquid helium temperature $T_{\mathrm{L}}$ in the main volume.
To reduce noise in $\dot{T}_{\mathrm{L}}\left(t\right)$, we averaged
the QTF data over 7 to 20 min intervals, depending on
the scatter of the data. Having $\dot{Q}_{\mathrm{tube}}$, we can then solve
$\kappa_{1}\left(T_{\mathrm{tube}}\right)$ from Eq.~\eqref{eq:tube-heat} as a function of the
channel temperature $T_{\mathrm{tube}}$. We have taken it to be the mean value between
$T_{\mathrm{V}}\left(t\right)$ and $T_{\mathrm{L}}\left(t\right)$,
when both are below the $T_{c}$, and the mean value between $T_{\mathrm{V}}\left(t\right)$
and $T_{c}$ when only the heat-exchanger volume is below the superfluid transition
temperature.

Figure \ref{fig:thermal-conductivity} shows the resulting thermal
conductivity, averaged across all analyzed precools and warm-ups.
The confidence bounds include the measurement spread, as well as 10\%
uncertainty in the channel dimension parameter $D$, and in the Kapitza
constants $r_{\mathrm{L}}$ and $r_{\mathrm{V}}$, and 5\% uncertainty
in the Kapitza exponents $p_{\mathrm{L}}$ and $p_{\mathrm{V}}$.
The solid black line indicates a fit to the experimental data of form
$G\left(T\right)=g_{1}\left(T\right)+\left(\mathcal{K}_{2}-g_{2}\left(T\right)\right)+g_{3}\left(T\right)$,
where $g_{i}=\mathcal{K}_{i}\exp\left[-\left(\left(T-T_{0,i}\right)/\sigma_{i}\right)^{2}\right]$
is a Gaussian function, with $\mathcal{K}_{i}$, $T_{0,i}$ and $\sigma_{i}$
listed in Table \ref{tab:fit-parameters}. Such an analytic form is handy for the model simulations.

The distinct features of our data are a plateau between $0.3T_{c}$
and $0.5T_{c}$, and a local maximum at $0.75T_{c}$. As we approach
the $T_{c}$ from below, the conductivity first decreases from the
local maximum value until about $0.85T_{c}$ after which it starts
to increase again until about $0.95T_{c}$. Data-analysis near the $T_c$ was challenging due to two things.
First, \textsuperscript{3}He usually undercooled slightly as we crossed
the $T_{c}$ from above, i.e., temperature of the liquid was already
below the $T_{c}$ but it was not yet in the superfluid state, and second,
our QTF calibration formula changed
at the $T_{c}$ from normal fluid viscosity dependent calibration \cite{Pentti_Rysti_Salmela}
to a phenomenological one. The combined effect of the changing calibration
and undercooling of the liquid causes a small artificial jump in the
temperature determined from the QTF frequency and width,
which results in a large apparent derivative $\dot{T}_{\mathrm{L}}\left(t\right)$
rendering our analysis inaccurate near the $T_{c}$. As a further
complication, at a certain range, we may have a situation, where
the main volume is in A-phase of the superfluid while the heat-exchanger volume
is already in the B-phase, and the A-B phase boundary can be somewhere
in the channel causing unpredictable behavior
in the determined thermal conductivity. With these issues acknowledged,
we conclude that our analysis gives reasonable thermal conductivity
data in the B-phase of \textsuperscript{3}He superfluid ($T<0.92T_{c}$).
\begin{wraptable}{O}{0.42\columnwidth}%
\vspace*{-0.03\textheight}

\begin{tabular}{|c|c|c|c|}
\hline 
$i$ & $\mathcal{K}_{i}/\kappa\left(T_{c}\right)$  & $T_{0,i}/T_{c}$  & $\sigma_{i}/T_{c}$ \tabularnewline
\hline 
\hline 
1 & 2.68 & 0.76 & 0.11\tabularnewline
\hline 
2 & 0.94 & 0 & 0.25\tabularnewline
\hline 
3 & 3.76 & 0.94 & 0.05\tabularnewline
\hline 
\end{tabular}

\caption{List of the parameters used in the multi-Gaussian fit of Fig.~\ref{fig:thermal-conductivity}\label{tab:fit-parameters}}

\vspace*{-0.03\textheight}
\end{wraptable}%

Johnson \emph{et al.} \cite{Johnson1975} reported anomalous thermal
resistance behavior in the A-phase near the melting pressure of \textsuperscript{3}He.
Their thermal resistance data, converted to thermal conductivity is
included in Fig.~\ref{fig:thermal-conductivity} for comparison, showing
roughly the same magnitude with our measurement. Their data is scaled by the normal fluid conductivity at the $T_c$ for each pressure ($0.047\,\mathrm{W/\left(K\,m\right)}$ at 2.00~MPa, $0.032\,\mathrm{W/\left(K\,m\right)}$ at 2.96~MPa \cite{Greywall1984}). Wellard \emph{et al.} \cite{Wellard1982} studied the conductivity of superfluid \textsuperscript{3}He at 2.1~MPa down to $0.3T_{c}$ by observing
a time delay of a heat pulse between two vibrating wires, that was
converted to diffusive conductivity by Einzel \cite{Einzel1984} as
normalized to the normal fluid conductivity. Feng \emph{et al.} \cite{Feng1993} determined area-scaled \textsuperscript{3}He thermal resistance 6.8 K$\,$m$^2$/W in 3.6~mm long channel at 0.4~mK, at the \textsuperscript{3}He crystallization pressure. That corresponds to thermal conductivity of about $0.02\kappa/\kappa\left(T_c\right)$, which is 5--16 times less than the conductivity determined from our measurement. Full correspondence
between all the described data sets is not to be expected due to the different
conditions in these experiments. 

Near the $T_{c}$ we need to take into account the hydrodynamic
thermal conductivity, which is given by \cite{London1948,Johnson1975}
\begin{equation}
\kappa_{\mathrm{h}}=\frac{d^{2}TS_{3}^{2}}{32\eta V^{2}},\label{eq:hydro_conductivity}
\end{equation}
where $d$ is the diameter of the liquid column, and $\eta$ its viscosity, while $V=26.76$~cm$^3$/mol \cite{Kollar200a} is the molar volume of \textsuperscript{3}He at 2.564~MPa.
We used the normalized viscosity data given by Einzel \cite{Einzel1984}
with the normal fluid viscosity given by Ref.~\cite{Pentti_Thermometry}.
It is evident that such a mechanism is needed to explain the efficient heat transfer close to the $T_{c}$,
but the hydrodynamic contribution alone falls off too quickly as
the temperature decreases. The sum of diffusive and hydrodynamic conductivities
shows fair resemblance to our data, while it still does not reproduce
the local maximum at $0.75T_{c}$.

\begin{figure}[t]
\includegraphics[width=1\columnwidth]{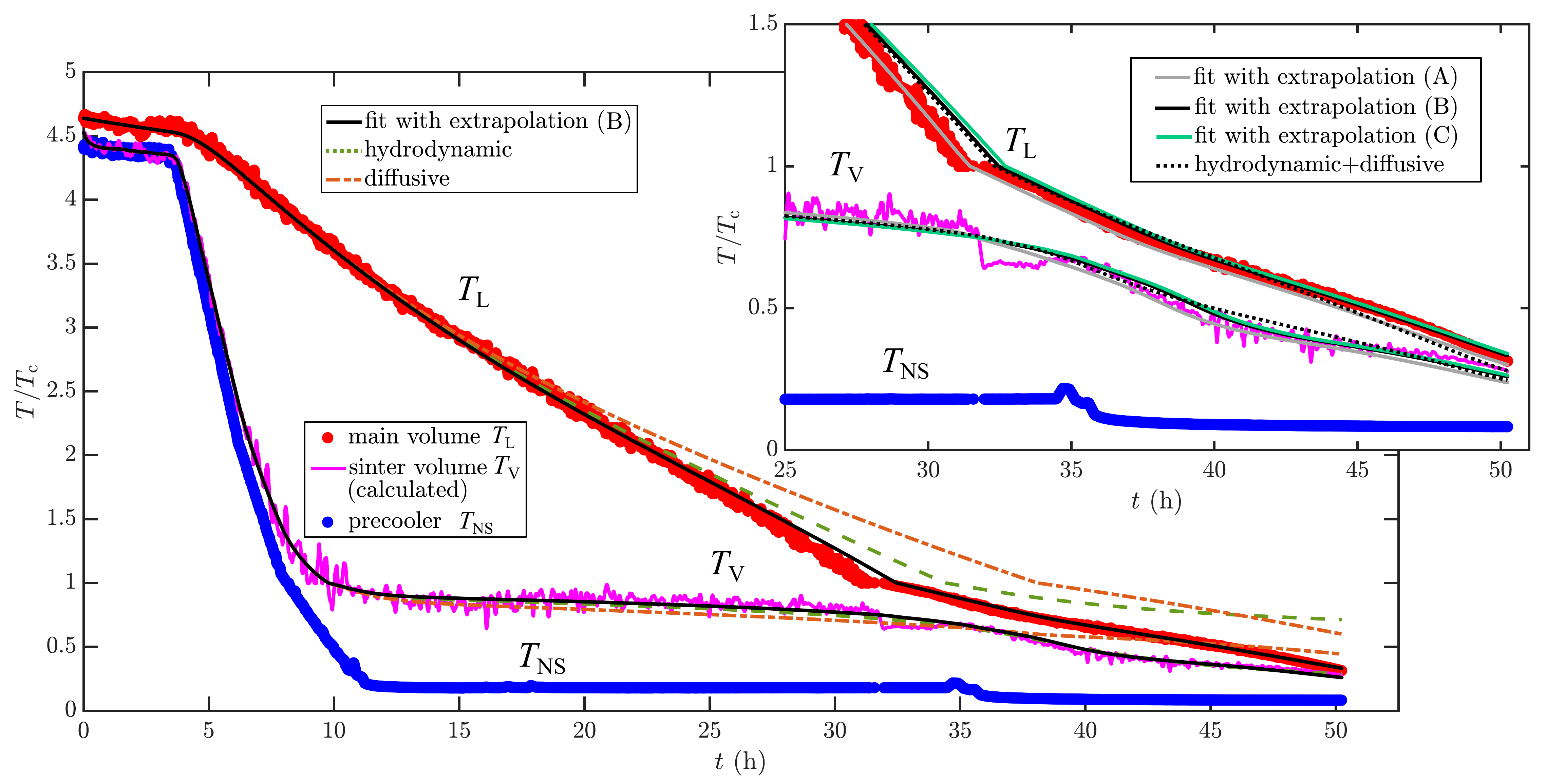}\caption{(color online) Example of the measured temperature of the main cell
volume $T_{\mathrm{L}}$ (red), the measured precooler temperature
$T_{\mathrm{NS}}$ (blue) and the computed heat-exchanger volume temperature
$T_{\mathrm{V}}$ (magenta) during a cooldown by the nuclear stage. The other curves
correspond to calculated main volume and heat-exchanger volume temperatures
using various superfluid \protect\textsuperscript{3}He conductivities
(see Fig.~\ref{fig:thermal-conductivity}). \emph{Main panel}: (solid
black) multi-Gaussian fit  $G\left(T\right)$ using the extrapolation (B), (dashed
green) hydrodynamic thermal conductivity, and (dash-dotted brown)
diffusive thermal conductivity. \emph{Inset: }(solid gray)  $G\left(T\right)$ with extrapolation (A), (solid black) extrapolation (B), (solid
green) extrapolation (C), and (dotted black) sum of hydrodynamic and
diffusive conductivities. The system had $570\,\mathrm{mmol}\,\left(\pm2\%\right)$
of \protect\textsuperscript{3}He in total; $344\,\mathrm{mmol\,\left(\pm2\%\right)}$
in the pure \protect\textsuperscript{3}He phase of the main volume,
$187\,\mathrm{mmol\,\left(\pm2\%\right)}$ in the heat-exchanger volume and
the connecting channel, $32\,\mathrm{mmol\,\left(\pm5\%\right)}$
in the mixture phase of the main volume, and at most $7\,\mathrm{mmol}$
stuck as mixture in the sinter. The amount of solid \protect\textsuperscript{4}He
was $2.98\,\mathrm{mol\,\left(\pm0.5\%\right)}$, while the external
heat leak was $40\,\mathrm{pW}$\label{fig:simulation}}
\end{figure}

Figure~\ref{fig:simulation} demonstrates how the main volume temperature
$T_{\mathrm{L}}$ is computationally reproduced using various \textsuperscript{3}He
thermal conductivities of Fig.~\ref{fig:thermal-conductivity}. It
also shows another crucial element of our analysis, the heat-exchanger volume
temperature $T_{\mathrm{V}}$ calculated from the measured main volume
temperature and the precooler temperature $T_{\mathrm{NS}}$. Scatter in the $T_{\mathrm{V}}$ data is due to the analysis procedure. We immediately
note from the main panel that neither diffusive nor hydrodynamic conductivity
alone can reproduce our observed data. The computed $T_{\mathrm{L}}$,
with either, starts to severely lag behind as the heat-exchanger volume goes
below the $T_{c}$.

The analysis is problematic in the \textsuperscript{3}He-A region
(from $0.92T_{c}$ to $T_{c}$), as our treatment is not accurate
there. Figure~\ref{fig:thermal-conductivity} shows three possible
extrapolations of the measured data, and the inset of Fig.~\ref{fig:simulation}
illustrates the resulting difference. The option (A) with linearly
increasing conductivity follows the measured data accurately above
the main volume $T_{c}$, but, from there downwards, it gives slightly
too low temperatures. The opposite is true for the options (B) with
constant conductivity, and (C) with the multi-Gaussian fit $G\left(T\right)$, as
both lag slightly behind the measured temperature above the $T_{c}$,
but give better correspondence below it. The combined diffusive and
hydrodynamic conductivities, based on data from the earlier publications, also reproduce the data
with decent accuracy, except at the lowest temperatures. This makes
sense as the combined conductivity is within the confidence bounds
of our measurements until $0.6T_{c}$, below which it stays too high
and thereby the computed main volume temperature would continue to
decrease more rapidly than the measured temperature.

\section{Conclusions\label{sec:Conclusions}}

We have determined the thermal conductivity of superfluid \textsuperscript{3}He-B
at the \textsuperscript{4}He crystallization pressure $2.564\,\mathrm{MPa}$
in a tubular channel connecting two volumes, larger of which contained
solid pure \textsuperscript{4}He, liquid saturated \textsuperscript{3}He\textendash \textsuperscript{4}He
mixture, and liquid pure \textsuperscript{3}He, while the smaller,
sinter-filled heat-exchanger volume, had solely pure \textsuperscript{3}He (with
possible traces of mixture within the sinter). The temperatures down
to $0.25T_{c}$ were covered during precooling the experimental cell
externally by a copper nuclear demagnetization cooler, while the temperatures
down to $0.1T_{c}$ were reached by utilizing the internal adiabatic
melting method and then observing the following warm-up. $0.1T_{c}$
was also the low temperature limit of our quartz oscillator thermometry. A handicap in our setup was that we could not directly measure the temperature of the heat-exchanger volume, but instead we had to resolve it from our computational model. To improve the setup an additional quartz tuning fork should be installed there.

At the onset of the B-phase $0.92T_{c}$, we observed thermal conductivity
4.3 times larger than that of normal fluid \textsuperscript{3}He
at the $T_{c}$. Then, as the temperature was lowered, the conductivity
showed a local minimum at $0.85T_{c}$ (2.5 relative units) followed
by a local maximum at $0.75T_{c}$ (3.5). Between $0.5T_{c}$ and $0.3T_{c}$ we observed a plateau at about 1 relative units, below which a monotonically
decreasing behavior was observed. At the plateau,
our data indicated $\sim60\%$ lower overall conductivity than the
value obtained from earlier studies \cite{Wellard1982,Einzel1984}.
We also showed that our measured temperature data was computationally
reproducible using the determined thermal conductivity, meaning that
the computational model can be used to estimate the lowest temperatures
reached by the adiabatic melting method, when the quartz oscillator thermometer
had become insensitive to temperature.

\begin{acknowledgements}
This work was supported by the Jenny and Antti Wihuri Foundation Grant
No. 00180313, and it utilized the facilities provided by Aalto University
at OtaNano - Low Temperature Laboratory.
\end{acknowledgements}

\section*{\textemdash \textemdash \textemdash \textemdash \textemdash \textemdash \textemdash{}}

\end{document}